\documentclass{acm_proc_article-sp}
\usepackage{epstopdf}
\usepackage{graphics}

\begin{document}

\conferenceinfo{}{Bloomberg Data for Good Exchange 2016, NY, USA}

\title{The Role of Rating and Loan Characteristics in Online Microfunding Behaviors}

\numberofauthors{3}
\author{
\alignauthor
Gaurav Paruthi\\
       \affaddr{ iSchool, University of Michigan}\\
       \affaddr{Ann Arbor, MI}\\
       \email{gparuthi@umich.edu}
\alignauthor
Enrique Frias-Martinez\\
       \affaddr{Telefonica Research}\\
       \affaddr{Madrid, Spain}\\
       \email{enrique.friasmartinez@telefonica.com}
\and
\alignauthor Vanessa Frias-Martinez\\
       \affaddr{iSchool, Univ. of Maryland}\\
       \affaddr{College Park, MD}\\
       \email{vfrias@umd.edu}
}

\maketitle






\section{Introduction}
{\it Microfinance institutions (MFIs)} like the Grameen Bank 
or the Banco do Nordeste in Brazil are non-profit organizations that 
give small loans to low-income borrowers, typically at low interest rates.
Their main aim is to contribute to the socioeconomic 
development of the regions where they operate while remaining financially sound. 

In recent years, there has been a significant growth in the number of online microlending
sites that connect individuals to small businesses led by low-income citizens.
This type of social lending platforms allow individuals from all over the world to explore large
online databases of businesses that require small loans to succeed; and citizens without access
to formal banking systems to borrow the money they need to carry out their projects.
Understanding lending and borrowing activity is critical to improve the way microfinance
services work. However, 
although there exists an important body of work regarding traditional (offline) microfinance,
the research in the area of online microlending is much more limited. 
Previous work in traditional microlending has covered various research questions including
the relationship between gender and loan reimbursement \cite{helena} or
 the impact that MFI ratings have on total investments and growth \cite{gutierrez,sufi}.
Nevertheless, there only exist a few studies that focus on the analysis of {\it online} microlending
platforms \cite{li,desai}. 
However, many aspects that have been typically addressed in traditional microlending studies such as
the role that MFI ratings or teams play in the lending process, have not been fully analyzed online. 

In this paper, we propose an in-depth study of lending behaviors in Kiva using a mix of quantitative and large-scale data mining techniques.
Kiva is a non-profit organization that offers an online platform to connect lenders with borrowers.
Their site, {\it kiva.org},  allows citizens to 
microlend small amounts of money to entrepreneurs (borrowers) from different countries. The borrowers are always 
affiliated with a Field Partner (FP)
which can be a microfinance institution (MFI) or other type of
local organization that has partnered with Kiva.
Field partners give loans to selected businesses based on their local knowledge regarding the country, the business sector including
agriculture, health or manufacture among others, and the borrower.
  Our objective is to understand the
relationship between lending activity and various features offered by the online platform.
Specifically, we focus on two  research questions: (i) the role that MFI ratings play in driving lending activity and 
(ii) the role that various loan features have in the lending behavior. 
The first question analyzes whether 
there exists a relationship between the MFI ratings -- that lenders can explore online -- and their lending volumes.
The second
research question attempts to understand if certain loan features -- available online at Kiva-- such
as the type of small business, the gender of the borrower, or the loan's country information might affect the way lenders lend.

We carry out our analysis with a dataset collected by Schaaf {\it et al.} in combination with the loan-lender data snapshots provided by Kiva and a set of socioeconomic indicators provided by the World Bank Open Data website.
The resulting four-month dataset contains over a million different lending 
actions ($1,217,627$), $47,790$ loans and $263,121$ unique lenders. 
%
We expect that this analysis will provide Kiva and other similar microlending platforms with 
findings and techniques to better cater to their lenders so as to facilitate and enhance lending activity.




\section{MFI Rating and lending activity}
\label{rq1method}

The first research question focuses on understanding the relationship between Field Partner 
ratings and lending activity. For that purpose, we analyze 
the relationship between the number of lending actions to specific loans
and the rating of the FPs associated to those loans. 

  Kiva statistics reveal that 
the distribution of loans per rating is not homogeneous with 
a large number of loans associated to Field Partners with ratings three ($>140K$) and zero ($>80K$),
 while other FP ratings have 
considerably smaller number of loans ($ \approx 50K$, on average).  
Similarly, 
a large number of Field Partners ($90$) have ratings between $2.5$ and $3.5$ 
while higher and lower
ratings are associated to fewer FPs. For example, there exist only 
$24$ FPs with ratings between $1$ and $2$ or $31$ between $4$ and $5$.
%
In order to account for these differences and to be able to compare and correlate lending actions across ratings, we 
define {\it normalized lending actions} as the total number of lending actions to a given rating divided
by the number of loans associated to FPs with that rating: 
$l.a_{r} = \frac{l.a_{r}}{loans_{r}}$ where $l.a_{r}$ represents the total number of individual lending actions to
FPs with rating $r$ and $loans_{r}$ the total number of loans offered by FPs with 
rating $r$.

\begin{figure}
\centering
\includegraphics[scale=0.18]{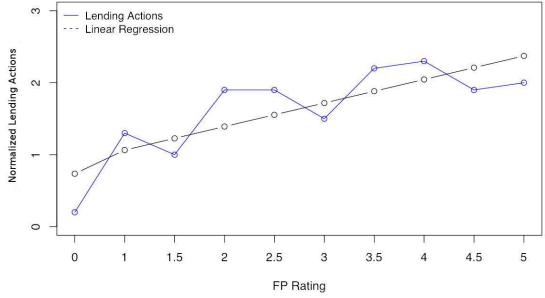}
\caption{Normalized lending actions per FP rating.}
\label{fig:FPratings}
\end{figure}

To test the relationship between lending activity and ratings, we compute the correlation coefficients between
the sum of all individual lending actions per rating ($l.a_{r}$) and the ratings themselves.
Figure \ref{fig:FPratings} shows the normalized activity for each rating value.
Since Kiva expresses the ratings as zero or a number between one and five in increments of $0.5$, the sample size
for the correlation analysis is only ten. Given that small size, we cannot guarantee normality or linearity. For
that reason we compute correlations for both parametric (Pearson's) and non-parametric (Spearman's rank) tests. 
Our analysis shows a strong positive correlation between the two with a correlation coefficient 
of $r(8)=0.78$ and p-value of $p=0.006$. 
Similarly, Spearman's rank produced a correlation coefficient of $\rho(8)=0.74$ (with $p=0.01$). 
Additionally, we also performed a linear regression on the ratings to see how predictive these are of the lending activity.
We obtained an $F(1,8)=10.24$ with $p=0.01$ and an adjusted $R^{2}=0.61$.
Thus, the tests determine that the trend in Figure \ref{fig:FPratings} 
approximately follows a monotonic linear trend.
This means that the higher the rating 
of a Field Partner, the larger the number of lending actions we observe.  In fact, lenders appear to 
be 
more prone to lend to loans managed by Field Partners with higher ratings. Similar results have been reported in 
one-to-one lending systems \cite{sufi}.

To better characterize the relationship between lending activity and ratings, 
we are also interested in understanding what type of lenders are more prone to lend
to highly rated Field Partners. 
Specifically, we seek lending patterns that exclusively characterize lenders whose 
lending activity is mostly focused on high FP ratings. 
For that purpose, we will first model each individual lender in our dataset with six different lending features:
(1) number of loans that the lender has lent money to, (2) invitee count, 
(3) number of days since she has been a member at Kiva, 
(4) average team size of the teams the lender is a member of, (5) number of distinct FP's the lender has lent to and 
(6) entropy of the lender's lending behavior.

Variables one to five are computed straightforward from the four-month dataset.
As for the lending entropy, we compute it as the Kolmogorov Complexity of the four-month 
lending actions' time series for each individual \cite{kolmogorov}.
Higher complexity values are associated
to burstier behaviors where lending patterns are harder to model as opposed to low complexity values
which we associate to more stable, planned lending behaviors {\it e.g.,} lenders that lend approximately
once every two weeks will have lower complexity values than those who lend more unpredictably.

\begin{figure}
\centering
\includegraphics[scale=0.5]{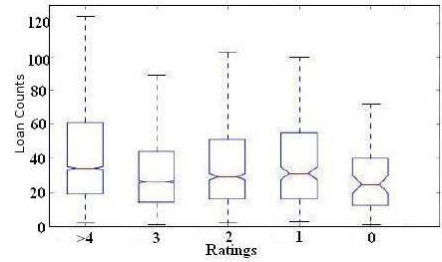}
\caption{Box plot for median individual loan counts versus ratings. Boxes represent the 1st and 3rd quartiles and values outside
the box are values within 1.5 the interquartile distance (1.5*Q3-Q1).}
\label{fig:loancount}
\end{figure}
\begin{figure}
\hskip 1.5cm
\includegraphics[scale=0.32]{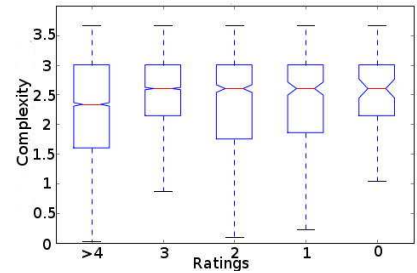}
\caption{Box plots for median individual complexity values that characterize lending patterns.} 
\label{fig:complexity}
\end{figure}

\begin{table*}
\centering
\small{
\begin{tabular}{|l|l|l|l|l|}
\hline
{\bf Name} & {\bf Code} & {\bf CC} & {\bf p-value} & {\bf q-value} \\ \hline
Urban population (in largest city) &EN.URB.LCTY.UR.ZS & $r(55)=0.332$ & 0.01 & 0.019 \\ \hline
Agriculture (added value) & NV.AGR.TOTL.KD & $r(51)=-0.23$ & 0.07 & 0.08\\ \hline
Strength of legal rights index (0=weak to 10=strong) &IC.LGL.CRED.XQ & $r(54)=-0.27$ &0.03 & 0.045 \\ \hline
Manufacturing, value added (\% of GDP) &NV.IND.MANF.ZS &$r(46)=-0.22$ &0.01 & 0.019 \\ \hline
Domestic credit provided by banking sector (\% of GDP) & FS.AST.DOMS.GD.ZS& $r(52)=-0.2$ &0.01 & 0.020 \\ \hline
Incidence of tuberculosis (per 100,000 people) &SH.TBS.INCD & $r(56)=-0.23$ &0.07 & 0.08\\ \hline
\end{tabular}
}
\caption{Pearson's correlations with FDR adjustments between normalized lending actions and WB indicators. }
\label{table:loanfeatures}
\end{table*}

To carry out this analysis, we characterize each individual in our dataset with the six lending features. 
For each feature and rating, we compute their median values and compare them.
For simplicity purposes, we consider five different ratings: zero, one ([1,2)), two ([2,3)), 
three ([3,4)) and four ([4,5]).
Our results reveal differences for two features:
(1) loan count and (6) complexity.
Figure \ref{fig:loancount} and Figure \ref{fig:complexity} show the mean loan count and the mean lending
complexity for lenders with a majority of lending actions on one of the five ratings, respectively.
We observe that lenders whose lending activity mostly focuses on FP's with ratings four or higher,
appear to lend to a larger number of loans while showing lower lending complexity than lenders that
focus their activity on lower ratings. 
In fact, it appears that lenders that concentrate on higher ratings might have more stable lending
behaviors probably implying regularly planned lending decisions. 
On the other hand, lenders whose majority lending actions are mostly focused
on FPs with lower ratings, appear to lend to fewer loans and their behaviors are far more complex,
which might reveal burstier, more impulsive behavior.
Without claiming causality, these results might suggest that offering more highly ranked Field Partners 
on Kiva's website could also potentially increase lending activity.
Additionally, given that planned lending decisions appear to be related to higher lending volumes, 
Kiva could offer planning tools to lenders such as lending calendars or lending reminders,  which
might also help to increase the lending activity of their users.

\section{Loan Features}

Our second research question seeks to understand the relationship between lending activity and features
that characterize a loan including: (1) country of the loan, (2) sector: agriculture, retail or health among others,
(3) size of the loan: individual or group-based and (4) gender of the borrowers in the loan. 
To eliminate the bias we normalize the number of 
lending actions to a loan feature (country, sector, gender or size) by dividing the total number of lending actions by 
the number loans associated to that feature as: 
 $l.a_{f} = \frac{l.a_{f}}{loans_{f}} \;with\; f\in\{c,s,g,z\}$
   where $f$ represents one of the loan features: the country $c$, 
the sector $s$, the gender $g$ or the size $z$ of the loan; $l.a_{f}$ the
number of lending actions to loans with feature $f$ and $loans_{f}$ the total
number of loans with that feature.

To evaluate how lending actions and the country of the loan relate to each other, 
we first characterize each borrower country by a set of over $1000$ socioeconomic indicators extracted from the
World's Bank Open Data website.
Next, we compute Pearson's correlations between the total number of 
lending actions ($l.a_{c}$) to each borrower country $c$ and the values for each socioeconomic variable. 
Given the large number of correlations performed (over $1000$), we need to adjust the p-values {\it i.e.,} control
for the Type I error (False Positives). Bonferroni correction is one of the most common approaches to adjust for multiple testing. However, given the large set of correlations that we perform, Bonferroni's p-values would be too stringent {\it e.g.,} for a Type I error rate of 0.01, it would require a $p >  0.00001$.
For that reason, we apply instead the False Discovery Rate (FDR) which controls the fraction of positive detections 
that are wrong. 
Specifically, we use the Benjamini-Yekutieli's FDR adjustment and report the correlation results together with 
their p-values and their q-values {\it i.e.,} percentage of false discoveries accepted for that test, also known as adjusted p-values \cite{benjamini2}.

\begin{table*}
\centering
\small{
\begin{tabular}{|l|l|l|l|l|}
\hline
{\bf Borrower Country} & {\bf Lender Country} & \bf{CC} & \bf{p} & {\bf q } \\ \hline
EN.URB.LCTY.UR.ZS & Low Average Interests on External Debt ($DT.INR.OFFT$) & $r(51)=0.38$ &0.004 &0.01\\ \hline
NV.AGR.TOTL.KD & Persistence to Last Grade Primary Education ($SE.PRM.PRSL.ZS$) & $r(49)=0.44$ & 0.002 & 0.008\\ \hline
IT.CEL.SETS & High Youth Literacy Rates ($SE.ADT.1524.LT.ZS$) &$r(53)=-0.35$  &0.001 &0.006\\ \hline
NV.IND.MANF.ZS & High Military Expenditure ($MS.MIL.XPND.GD.ZS$) & $r(49)=-0.45$ & 0.001 &0.006 \\ \hline
SH.TBS.INCD & Low Prevalence of overweight children ($SH.STA.OWGH.ZS$) & $r(46)=0.38$ &0.003 &0.008\\ \hline
\end{tabular}
}
\caption{Pearson's correlations with FDR adjustments between lending activity of lender countries characterized by low, medium or high socioeconomic indicators and 
   borrower countries socioeconomic values'.}
\label{table:loancountries}
\end{table*}

Table \ref{table:loanfeatures} shows some of the most relevant findings (the majority with $q \leq 0.05$).
We observe a significant positive correlation between a country's urban population and the lending activity it receives
(recall that lending activity is normalized by the number of loans in the country).
It appears that countries with a larger urban population have a higher probability of benefiting from lending than countries that are more rural.
To promote more heterogeneous lending activity, Kiva could explore putting online {\it lending recommendations} 
to drive lending activity towards countries that benefit the least at each moment in time.  
Interestingly, we also observe a negative correlation
between lending activity towards a country and its agricultural added value. 
This shows that Kiva lenders are lending more to countries whose agricultural production is poor and in need
(the smaller the production, the larger the lending activity).

Other correlations indicate that larger lending activity is associated to countries with less manufacturing which is indirectly related to job creation. In fact,
the lack of manufacturing industries can negatively impact the creation of jobs. 
As a result, Kiva appears to be successfully driving
lending activity towards countries where job creation is harder to achieve. On the other hand, lending activity is 
negatively
correlated to the strength of legal rights in the country, which might
reveal a lending activity focused on supporting development and indirectly the improvement of freedom and rights  
in borrower countries.

To understand better lending actions, it is important to realize that lenders 
can also be influenced by the socioeconomic conditions of their own countries.
In an attempt to disentangle which factors play a role, we also analyze the relationship
between borrower countries and the lending activity of lender countries characterized by their socioeconomic indicators.
For that purpose, we compute the total number of lending actions per
lender country to each borrower country. Next, we characterize each lender country with its socioeconomic
indicators extracted from the World's Bank Open Data. For each indicator, we create
three groups of lender countries depending on whether the country has a low, medium or high value for that indicator, 
and compute their total lending activity to each borrower country. 
This will allow us to refer to the lending activity of, for example, {\it lender countries that have a low GDP} or
{\it lender countries with high mobile cellular subscriptions}.
Next, for each lender indicator (GDP, inflation,...) and group (low, medium or high), we  
compute Pearson's correlations (adjusted with FDR) between their lending activities to each borrower country and the values for each socioeconomic indicator from the borrower countries. 
These tests might reveal important relationships between groups of lender countries and borrower countries 
being able to draw statements such as 
{\it the lower the GDP of 
   the borrower country, the larger lending activity they attract from countries with high GDP}.

Table \ref{table:loancountries} shows some of
the most relevant correlations between the indicators of the borrower countries previously discussed and 
the lending activity they receive from countries with certain low, medium or high socioeconomic values.
We observe a positive correlation between lender countries that have low average interests on external 
debt and borrower countries 
with large urban populations. As discussed earlier, borrower countries with large urban populations seem to 
receive more lending activity which they appear to be getting from lender countries that are not strangled by their external debt payments.
We also observe that lender countries where most citizens finish primary education 
have a lending activity that is positively correlated to 
the borrower's agricultural production ($NV.AGR.TOTL.KD$). This implies that the little lending 
activity that borrower countries with large agricultural production
manage to bring in (as shown in Table \ref{table:loanfeatures}) is mostly from lender countries with high education
levels. 
In terms of mobile penetration ($IT.CEL.SETS$), Table \ref{table:loanfeatures} showed a tendency to lend more to countries with
low penetration rates, and Table \ref{table:loancountries} shows that it is mostly countries with high youth literacy rates,
the ones who generate that lending activity. 
We also observe that countries with high military expenditure ($MS.MIL.XPND.
 GD.ZS$) focus their lending 
activity on borrower countries
with low manufacturing rates ($NV.IND.MANF.ZS$).
Additionally, lending actions to countries with high incidence of tuberculosis appear to be mostly driven by countries with 
low prevalence of overweight children ($SH.STA.OWGH.ZS$). We posit that countries that are aware of the importance of
health related issues might be focusing on lending to countries who could improve their health status.
To summarize, it is fair to say that the general trend is for countries with higher rates of educated citizens and larger economic activity to be more
prone to lend, which is a feature that has also been found in official development assistance \cite{desai}.

Finally, to analyze the relationship between lending activity and the sector of the loan,
the size of the loan and the gender, we take a different approach to account for the discrete nature of the variables.
We compute the median number of normalized
lending actions and its standard deviations for: each type of loan sector (health, agriculture, retail, etc.);
 each group size range(1, [2-10], [10-20] and [20-48]) and for each gender (female, female or both for loans that go to a group of borrowers rather than an individual).
In terms of sectors, we
observe a larger median number of normalized lending actions in the retail sector with 
(M=3.2, IQR=6.2), followed
by the agricultural (M=3.1,IQR=6.2) and food (M=2.9,
IQR=6.1) sectors, where $M$ represents the median
and $IQR$ the interquartile range. 
The other sectors
showed considerably lower values, although these differences were
not statistically significant.
This shows that lenders appear to favor the retail sector which is probably far more
present in urban than rural settings. 
In terms of group size, we observe that individuals appear to focus their lending activity 
on loans which are borrowed by one person (M=2.3, IQR=3) or group loans
of up to ten individuals (M=4.1, IQR=6). Larger loans are not as favored which is also coherent with
the findings reported in \cite{ghatak,owusu}. Finally, gender presents a slight minimal advantage
for female loans (M=2.4, IQR=3) versus male (M=2.1, IQR=3), but nothing conclusive, although similar results
have been reported in \cite{li,pitt,mayoux}.
These lending patterns might be a result of individual preferences or
rather a consequence of the way Kiva presents the information on their website.

\section{Conclusions}
\label{conclusions}
We have presented a large-scale analysis of the role that various features might play
on online microlending environments. 
Our results show that lenders appear to favor highly rated Field Partners that manage to drive more
lending activity. Additionally, we have observed that lenders seem to lend to loans in
sectors that are often times aligned with official aid donors. 
We believe that our work provides a better understanding of online microlending behaviors as well
as a set of suggestions to improve the services that microfinance platforms currently offer to their lenders and borrowers.

%
%
%

\bibliographystyle{plain}
\bibliography{dev}

\end{document}